\documentclass[aps,prd,preprint]{revtex4-1}
\usepackage {graphicx}
\usepackage[toc,page]{appendix}
\usepackage{amsmath}
\usepackage{tikz}
\newcommand{\be}{\begin{equation}}
\newcommand{\ee}{\end{equation}}
\newcommand{\bea}{\begin{eqnarray}}
\newcommand{\eea}{\end{eqnarray}}
\begin{document}
\title{A mean curvature flow method for numerical cosmology}
\author{Matthew Doniere}
\email{mdoniere@oakland.edu}
\affiliation{Dept. of Physics, Oakland University, Rochester, MI 48309, USA}
\author{David Garfinkle}
\email{garfinkl@oakland.edu}
\affiliation{Dept. of Physics, Oakland University, Rochester, MI 48309, USA}

\date{\today}

\begin{abstract}
We provide a mean curvature flow method for numerical cosmology and test it on cases of inhomogenous inflation.  The results show (in a proof of concept way) that the method can handle even large inhomogeneities that result from different regions exiting inflation at different times.

\end{abstract}


\maketitle

\section{Introduction}

In recent years there have been several numerical studies of inhomogeneous expanding cosmologies.\cite{east1,east2,clough1,clough2,clough3,joana,meannapaul}  Ultimately the goal of these simulations is to use a wide enough class of initial data and to evolve long enough to determine whether inflation occurs generically. These simulations use several different slicing conditions, with some using generalized harmonic coordinates, others using the puncture gauge, and still others using constant mean curvature (CMC) slicing.  While any of these slicing conditions can evolve for some time, it is not clear whether (in all cases of interest) they can evolve long enough to extract all relevant physics.  

This issue of long time evolution is addessed from the mathematical side in a recent paper by Wang and Senatore\cite{senatore} which uses mean curvature flow to study inflationary cosmology.  In this paper the authors show the long time existence of their slicing and the asymptotic behavior of the spacetime.  Mean curvature flow has been extensively studied by pure mathematicians, but has (so far) seen comparitively little application in physics.

We are thus motivated to try mean curvature flow slicing as a numerical method to study expanding cosmologies.  The variables, equations of motion, and numerical methods used are described in section~\ref{equations}.  Our results are presented in section~\ref{results}.  Conclusions are given in section~\ref{Conclude}.

\section{Equations of motion}
\label{equations}

The spacetime is described in terms of a coordinate system $(t,{x^i})$ and a tetrad $({{\bf e}_0 ^a},{{\bf e}_\alpha ^a})$, where both $i$ and $\alpha$ go from 1 to 3. 
We choose ${\bf e}_0$ to be hypersurface orthogonal with the relation between tetrad and coordinates of the form
${{\bf e}_0}={N^{-1}}{\partial _t}$ and ${{\bf e}_\alpha} = {{e_\alpha}^i}{\partial _i}$.  Here $N$ is the lapse and the shift is chosen to be zero.  Note that this means that for any quantity $F$ we have ${\partial _t}F = N {e_0}(F)$.  Choose the spatial triad to be Fermi propagated along the integral curves of ${\bf e}_0$.
 
The commutators of the tetrad components are decomposed as follows:
\bea
[{{\bf e}_0} , {{\bf e}_\alpha} ] &=& {{\dot u}_\alpha} {{\bf e}_0} \; - \; ( H {{\delta _\alpha}^\beta} + {{\sigma _\alpha}^\beta}){{\bf e}_\beta}
\\
{[{{\bf e}_\alpha} , {{\bf e}_\beta} ]} &=& \left ( 2 {a_{[\alpha}}{{\delta _{\beta ]}}^\gamma} \; + \; {\epsilon _{\alpha \beta \delta}}{n^{\delta \gamma}} \right ) {{\bf e}_\gamma}
\eea
where $n^{\alpha \beta}$ is symmetric, and $\sigma^{\alpha \beta}$ is symmetric and trace-free.  In physical terms, $H$ is one third of the mean curvature and is therefore equal to the Hubble constant when the spacetime is Friedmann-Lemaitre-Robertson-Walker (FLRW).  The shear $\sigma_{\alpha \beta}$ gives the extent to which different directions are expanding at different rates.  The quantity ${\dot u}_\alpha$
is not an independent variable, but is given in terms of the lapse by ${{\dot u}_\alpha} = {N^{-1}}{{\bf e}_\alpha}N$.  

The matter is a scalar field $\phi$ with potential $V$.  In order to obtain evolution equations for the matter variables that are first order in space and time, we define the quantities $P$ and $S_\alpha$ by $P \equiv {{\bf e}_0}(\phi )$ and 
$ {S_\alpha} \equiv {{\bf e}_\alpha} (\phi )$.

Mean curvature flow means that the surfaces of constant time evolve by flowing along their normal vector an amount equal to the mean curvature.  In terms of our variables, this means that the lapse $N$ is given by $N=3H$. Note that this means that 
${{\dot u}_\alpha} = {H^{-1}} {e_\alpha}H$. 

The evolution equations for the tetrad and matter quantities are as follows:
\bea
{\partial_t}  {{e_\alpha}^i} &=& - N ( H {{\delta _\alpha}^\beta} + {{\sigma _\alpha}^\beta}) 
{{e_\beta}^i}
\label{dte}
\\
{\partial_t}  H  &=& {e^\alpha}{e_\alpha} H \; - \; 2 {a^\alpha}{e_\alpha} H \; + \; N \left [  - \; {H^2} \; 
\; - \; {\textstyle {\frac 1 3}} {\sigma _{\alpha \beta}}{\sigma ^{\alpha \beta}} \; - \; 
{\textstyle {\frac 1 3}} ({P^2} - V) \right ]
\label{dtH}
\\
{\partial _t}  {a_\alpha}  &=& N \biggl [3 {{\bf e}_\alpha} ( H ) \; - \; {\textstyle {\frac 3 2}} 
{{\bf e}_\beta} ( {{\sigma _\alpha}^\beta}) \; - \; H ( {{\dot u}_\alpha} + {a_\alpha} ) 
\; + \; {{\sigma _\alpha}^\beta} \left ( {\textstyle {\frac 1 2}} {{\dot u}_\beta} + 5 {a_\beta} \right ) 
\nonumber
\\
 &+& \; 2 {\epsilon _{\alpha \beta \gamma}}{n^{\beta \delta}}{{\sigma_\delta}^\gamma} \; + \; 2 P {S_\alpha}
 \biggr ]
 \label{dta}
\\
{\partial _t} {n^{\alpha \beta}} &=& N \left [- {\epsilon ^{\gamma \delta ( \alpha}} {{\bf e}_\gamma} ( {{\sigma _\delta}^{\beta )}} ) \; - \; H {n^{\alpha \beta}} \; 
+ \; 2 {{n^{(\alpha}}_\lambda} 
{\sigma ^{\beta ) \lambda}} \; - \; {\epsilon ^{\gamma \delta (\alpha}} {{\dot u}_\gamma}
{{\sigma _\delta}^{\beta )}} \right ]
\label{dtn}
\\
{\partial_t} {\sigma _{\alpha \beta}}  &=& 3 {{\bf e}_{<\alpha}}{{\bf e}_{\beta >}} H \; + \;  N \biggl [ - \; {{\bf e}_{<\alpha}} ( {a_{\beta >}} ) \; - \; 3 H {\sigma _{\alpha \beta}}  \; + \; {a_{<\alpha}} {{\dot u}_{\beta >}} 
+ \; {\epsilon _{\gamma \delta ( \alpha}} {{\bf e}^\gamma} ( {{n_{\beta )}}^\delta} ) 
\nonumber
\\
&+& \;  {\epsilon _{\gamma \delta ( \alpha}} {{n_{\beta )}}^\delta} ( {{\dot u}^\gamma} - 2 {a^\gamma} ) \; - \; 2 {{n_{<\alpha}}^\gamma}{n_{\beta > \gamma}} + n {n_{<\alpha \beta >}} 
\; + \; {S_{<\alpha}}{S_{\beta >}} \biggr ]
\label{dtsigma}
\\
{\partial_t} \phi  &=& N P
\label{dtphi}
\\
{\partial_t} {S_\alpha}  &=& N \left [{{\bf e}_\alpha} ( P ) \; + \; P {{\dot u}_\alpha} \; - \; H {S_\alpha} \; - \; {{\sigma _\alpha}^\beta}{S_\beta} \right ]
\label{dtS}
\\
{\partial_t} P &=& N \left [ {{\bf e}^\alpha} ( {S_\alpha} ) \; - \; 3 H P \; + \; {S_\alpha} 
({{\dot u}^\alpha} - 2 {a^\alpha} ) \; - \; {\frac {dV} {d \phi}} \right ]
\label{dtP}
\eea

These variables are also subject to the vanishing of the following constraint quantities: 
\bea
{{\cal C}_{\rm com}} &=& {\epsilon^{\alpha \beta \lambda}} \left ( {e_\alpha}({{e_\beta}^i}) - {a_\alpha} {{e_\beta}^i} \right ) - {n^{\lambda \gamma}} {{e_\gamma }^i}
\label{cnstrcom}
\\
{{\cal C}_{u2}} &=& {\epsilon^{\alpha \beta \lambda}} \left ( {e_\beta}({{\dot u}_\alpha}) + {a_\alpha} {{\dot u}_\beta} \right ) + {n^{\lambda \gamma}} {{\dot u}_\gamma}
\label{cnstru2}
\\
{{\cal C}_J} &=& {e_\alpha}({n^{\alpha \delta}}) + {\epsilon^{\alpha \beta \delta}}
{e_\alpha}({a_\beta}) - 2 {a_\alpha}{n^{\alpha \delta}}
\label{cnstrJ}
\\
{{\cal C}_C} &=& {e_\beta} ( {{\sigma_\alpha}^\beta}) - 2 {e_\alpha}(H) - 3 {{\sigma_\alpha}^\beta}{a_\beta} - {\epsilon_{\alpha \beta \gamma}}{n^{\beta \delta}}{{\sigma_\delta}^\gamma} - P {S_\alpha}
\label{cnstrC}
\\
{{\cal C}_G} &=& 4 {e^\alpha} ({a_\alpha}) + 6 {H^2} - 6 {a^\alpha}{a_\alpha} - {n^{\alpha \beta}}{n_{\alpha \beta}} + {\textstyle {\frac 1 2}}{n^2} - {\sigma _{\alpha \beta}}{\sigma ^{\alpha \beta}} 
\nonumber
\\
&-& \left ( {P^2} + {S^\alpha}{S_\alpha} + 2 V \right )
\label{cnstrG}
\\
{{\cal C}_S} &=& {S_\alpha} - {e_\alpha}(\phi)
\label{cnstrS}
\eea
Initial data are chosen to solve the constraints of eqns. (\ref{cnstrcom}-\ref{cnstrS}) which are then preserved (to within numerical truncation error) under evolution. Preservation of the constraints up to truncation error is used as a code test and as a test that the resolution is adequate. The data are evolved using eqns. (\ref{dte}-\ref{dtP}).  Here eqn. (\ref{dtH}) is parabolic, and to obtain a mixed hyperbolic-parabolic system we add a multiple of eqn. (\ref{cnstrC}) to the right hand side of eqn. (\ref{dta}).

The hyperbolic equations are evolved using the iterated Crank-Nicholson method with the time step proportional to the space step as required by the Courant condition.  Numerical evolution of parabolic systems using explicit methods is usually slow because it requires a time step proportional to the square of the space step.  Instead we use an implicit method to treat eqn. (\ref{dtH}), which allows us to use the same time step as for the hyperbolic equations.

\section{Results}
\label{results}

To perform simulations quickly and with high resolution, we treat spacetimes with two spatial symmetries.  We use Cartesian coordinates $(x,y,z)$ and have dependence only on $x$.  We use periodic boundary conditions with $0 \le x \le 2 \pi$ with $0$ and $2\pi$ identified.

Initial data are found using the York method \cite{York}. That is, we write the initial data in terms of a freely specifiable piece and an unknown conformal factor which we solve for numerically.  The initial data for the metric variables are the following: 
\bea
H &=& {h_0}
\\
{{e_\alpha}^i} &=& {\psi ^{-2}} {{\delta _\alpha}^i}
\\
{a_\alpha} &=& - 2 {\psi ^{-1}} {{e_\alpha}^i} {\partial _i}\psi
\\
{n_{\alpha \beta}} &=& 0
\\
{\sigma _{\alpha \beta}} &=& {\psi ^{-6}} {Z_{\alpha \beta}}
\eea
Here $h_0$ is a constant, which means that our initial data surface is a constant mean curvature surface.
 
The initial data for the matter variables are as follows: $P={\psi^{-6}}Q$ and ${S_\alpha}={{\bf e}_\alpha}\phi$ where $Q$ and $\phi$ are given by 
\bea
Q &=& {p_0} + {f_0} \cos x
\\
\phi &=& {\phi_0} + {f_1} \cos x
\eea
where ${p_0},\, {f_0}, \, {\phi_0},$ and $f_1$ are constants.

For consistency with the momentum constraint (eqn. (\ref{cnstrC})) we pick $Z_{\alpha \beta}$ to be
\be
{Z_{\alpha \beta}} = {\rm diag}({a_{11}}, \lambda {a_{11}}, -(1+\lambda){a_{11}})
\ee
where $\lambda$ is a constant and $a_{11}$ is given by
\be
{a_{11}} = {f_1} \left ( {p_0} \cos x + {\textstyle {\frac 1 4}} {f_0} \cos 2x \right )
\ee 

The Hamiltonian constraint (eqn. (\ref{cnstrG}) then becomes the following elliptic equation for $\psi$
\be
{\partial ^i}{\partial _i} \psi \; + \;  {\textstyle {\frac 1 4}} (V(\phi) - 3 {H^2}) {\psi ^5} \; + \; 
{\textstyle {\frac 1 8}} \left ( {Q^2} +  {Z^{ij}}{Z_{ij}} \right ) {\psi ^{-7}} \; + \; ({\partial ^i}\phi {\partial _i}\phi ) \psi = 0
\ee
which we solve numerically.

\begin{figure}
\centering
\includegraphics[width=0.8\textwidth]{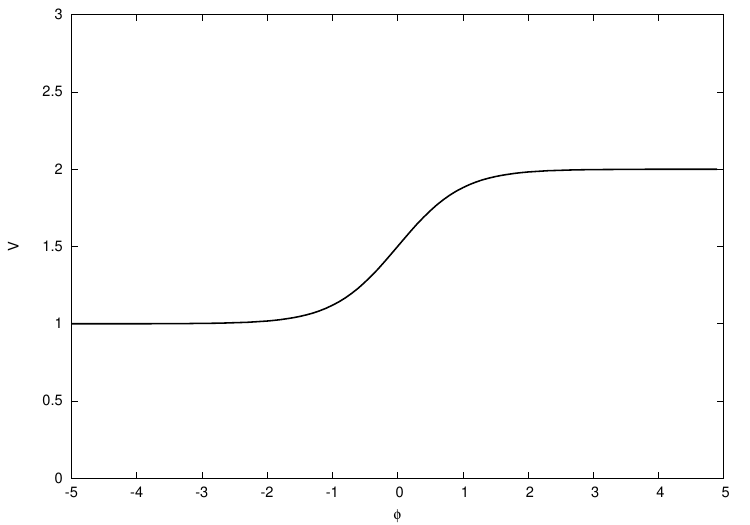}
\caption{$V$ vs. $\phi$ for ${\Lambda_1}=1, {\Lambda_2}=2$ and $c=1$. } 
\label{Vfig}
\end{figure}

One of the conditions for the theorems of \cite{senatore} is a potential $V$ satisfying $0 < {\Lambda _1} \le V \le {\Lambda _2}$ where $\Lambda _1$ and $\Lambda _2$ are constants.  In order to investigate this case, we choose a potential of the form
\be
V(\phi) = {\frac {{\Lambda _1}{e^{-\phi/c}} + {\Lambda _2}{e^{\phi/c}} }
{{e^{-\phi/c}} + {e^{\phi/c}} }}
\ee
where ${\Lambda _1}, \, {\Lambda _2},$ and $c$ are constants.  A plot of the potential with parameters ${\Lambda_1}=1, {\Lambda_2}=2, c=1$ is shown in figure \ref{Vfig}.  Note that this potential has two plateaus: the upper one at $\Lambda_2$ and the lower one at $\Lambda_1$.  We perform runs with $\Lambda _1$ and $\Lambda _2$ greater than zero to compare to the results of \cite{senatore}.  However, as noted in \cite{senatore} their conditions are only designed to model the onset of inflation.  In particular, with ${\Lambda _1} > 0$ one cannot model the exit from inflation.  In order to treat this case too, we also perform simulations with ${\Lambda _1}=0$.

Potentials of this form are chosen in this preliminary study for purposes of comparison with the results of \cite{senatore}.  However, our mean curvature flow method is compatible with any potential, including more commonly used potentials like ${m^2}{\phi^2}$.

\begin{figure}
\centering
\includegraphics[width=0.8\textwidth]{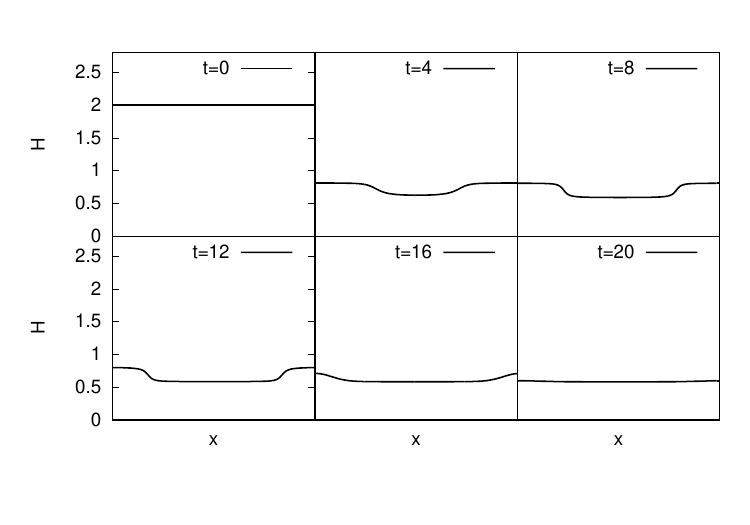}
\caption{$H$ vs. $x$ for $t=0, 4, 8, 12, 16$ and $20$. The parameters for the potential are ${\Lambda_1}=1, {\Lambda_2}=2, c=1$.  The parameters for the initial data are ${h_0}=2, {p_0}=5, {f_0}=0.1, {\phi_0}=0.5, {f_1}=0.8, \lambda=0.5$.} 
\label{Hfig1}
\end{figure}
 
\begin{figure}
\centering
\includegraphics[width=0.8\textwidth]{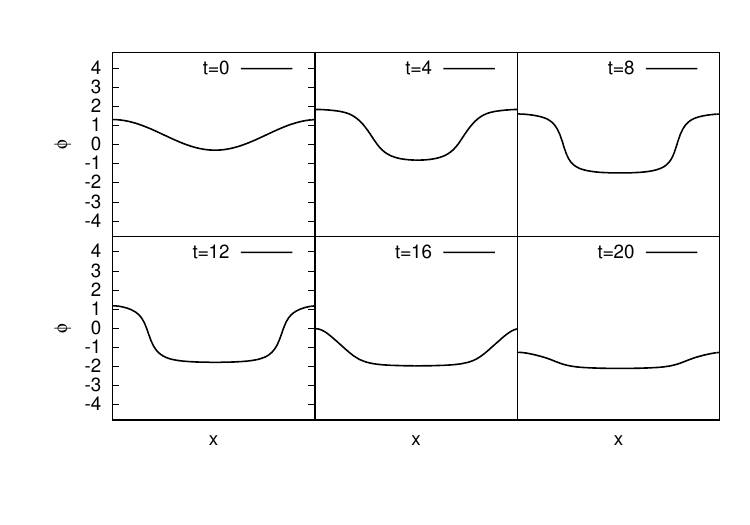}
\caption{$\phi$ vs. $x$ for $t=0, 4, 8, 12, 16$ and $20$. The parameters for the potential are ${\Lambda_1}=1, {\Lambda_2}=2, c=1$.  The parameters for the initial data are ${h_0}=2, {p_0}=5, {f_0}=0.1, {\phi_0}=0.5, {f_1}=0.8, \lambda=0.5$.} 
\label{phifig1}
\end{figure}

Results for a simulation using the potential of figure \ref{Vfig} are shown in figures \ref{Hfig1} and \ref{phifig1}.  Here figure \ref{Hfig1} shows the time development of $H$ while figure \ref{phifig1} shows the time development of $\phi$.  Since the coordinate $x$ has $0$ and $2\pi$ identified, this means that the left hand side of each panel of each graph is identified with the right hand side.  Note in figure \ref{Hfig1} that at intermediate times two regions develop with two different values of $H$, while by the end there is a uniform value of $H$ corresponding to what was the lower value at intermediate times.  This behavior can be understood by looking at the corresponding panels of figure \ref{phifig1}: at intermediate times one region has the scalar field $\phi$ at the top plateau of the potential, while the other region has $\phi$ at the bottom plateau of the potential.  By the final time of the simulation, $\phi$ is at the bottom plateau everywhere.

In a sense, these two regions are present from the begining, since the field starts out at the top plateau in one region and the bottom plateau in another region. From this point of view, it may seem odd that $H$ is uniform in the initial data.  However, constant $H$ is a requirement of the York method, since it allows the momentum constraint equation to decouple from the Hamiltionian constraint equation and thus be solved independently. 

\begin{figure}
\centering
\includegraphics[width=0.8\textwidth]{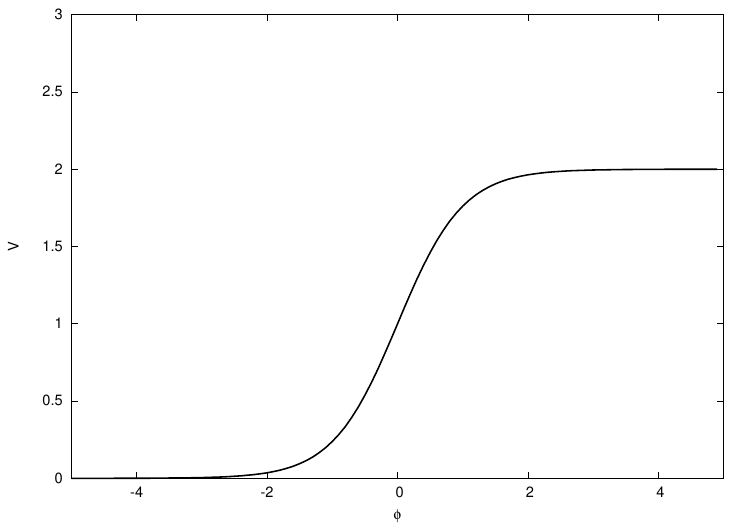}
\caption{$V$ vs. $\phi$ for ${\Lambda_1}=0, {\Lambda_2}=2$ and $c=1$. } 
\label{Vfig2}
\end{figure}

\begin{figure}
\centering
\includegraphics[width=0.8\textwidth]{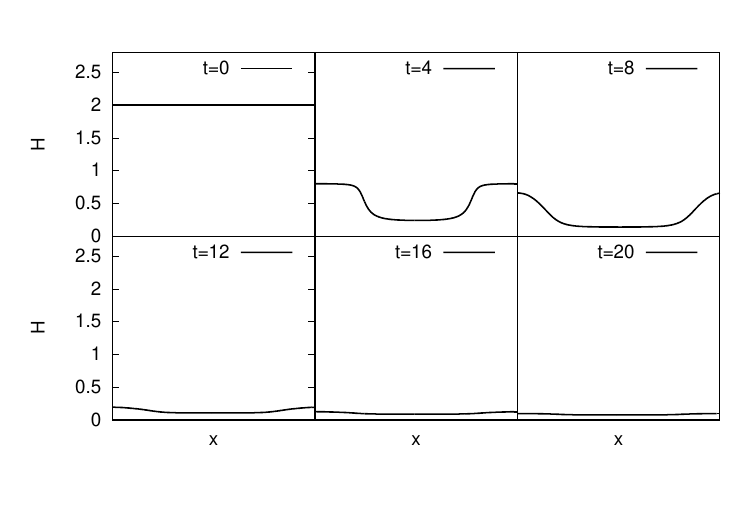}
\caption{$H$ vs. $x$ for $t=0, 4, 8, 12, 16$ and $20$. The parameters for the potential are ${\Lambda_1}=0, {\Lambda_2}=2, c=1$.  The parameters for the initial data are ${h_0}=2, {p_0}=5, {f_0}=0.1, {\phi_0}=0.5, {f_1}=0.8, \lambda=0.5$.} 
\label{Hfig2}
\end{figure}

\begin{figure}
\centering
\includegraphics[width=0.8\textwidth]{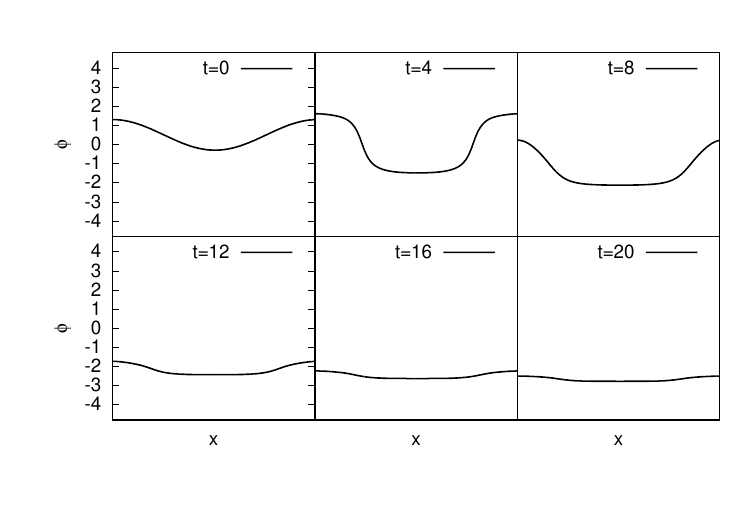}
\caption{$\phi$ vs. $x$ for $t=0, 4, 8, 12, 16$ and $20$. The parameters for the potential are ${\Lambda_1}=0, {\Lambda_2}=2, c=1$.  The parameters for the initial data are ${h_0}=2, {p_0}=5, {f_0}=0.1, {\phi_0}=0.5, {f_1}=0.8, \lambda=0.5$.} 
\label{phifig2}
\end{figure}

In order to model both inflation and the exit from inflation, we perform simulations with the potential given in figure \ref{Vfig2}, which has ${\Lambda_1}=0$ and ${\Lambda_2}=2$.  Results for a simulation with this potential is shown in figures \ref{Hfig2} and \ref{phifig2}.  Other than the change in the potential, this simulation uses the same parameters for initial data as the simulation presented in figures \ref{Hfig1} and \ref{phifig1}.  Note that the overall behavior is very similar to that of the previous simulation: figure \ref{Hfig2} shows that at intermediate times there are two regions with different values of $H$, while by the end of the simulation $H$ has become fairly uniform and is evolving towards zero.  Once again, this behavior can be understood by looking at the behavior of $\phi$ given in figure \ref{phifig2}: at intermediate times there are two regions, one with $\phi$ at the top plateau of the potential, while the other has $\phi$ at the bottom plateau of the potential.  By the end of the simulation, $\phi$ is at the bottom plateau everywhere.

\begin{figure}
\centering
\includegraphics[width=0.8\textwidth]{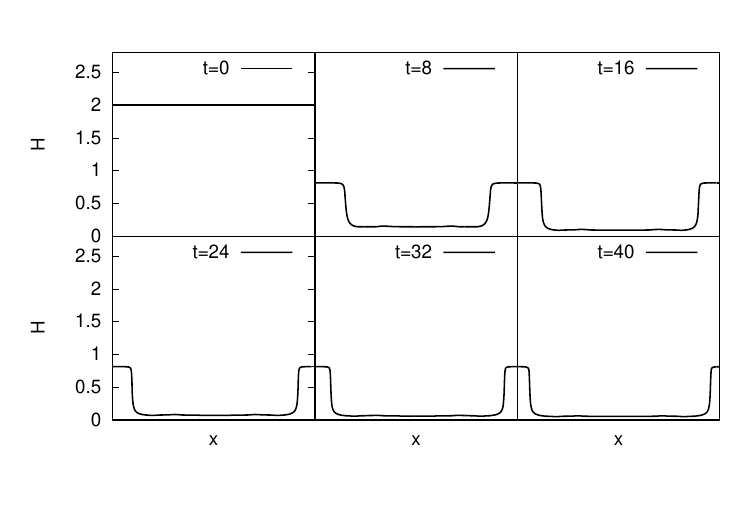}
\caption{$H$ vs. $x$ for $t=0, 8, 16, 24, 32$ and $40$. The parameters for the potential are ${\Lambda_1}=0, {\Lambda_2}=2, c=1$.  The parameters for the initial data are ${h_0}=2, {p_0}=1, {f_0}=0, {\phi_0}=0, {f_1}=3, \lambda=0.5$.} 
\label{Hfig3}
\end{figure}

\begin{figure}
\centering
\includegraphics[width=0.8\textwidth]{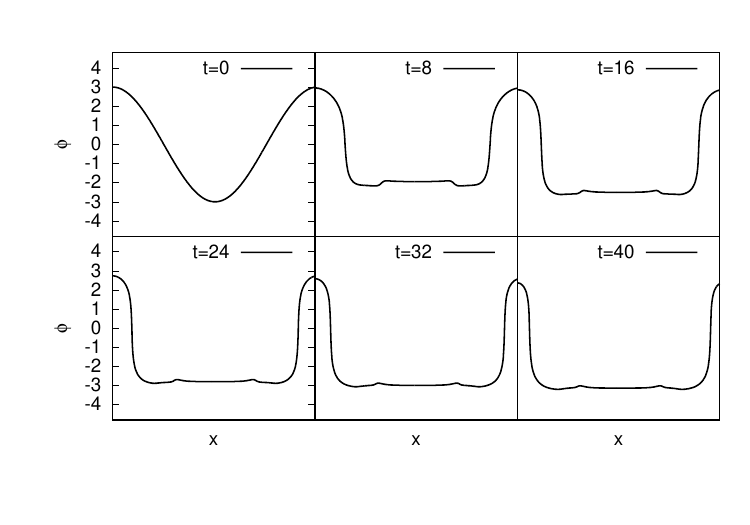}
\caption{$\phi$ vs. $x$ for $t=0, 8, 16, 24, 32$ and $40$. The parameters for the potential are ${\Lambda_1}=0, {\Lambda_2}=2, c=1$.  The parameters for the initial data are ${h_0}=2, {p_0}=1, {f_0}=0, {\phi_0}=0, {f_1}=3, \lambda=0.5$.} 
\label{phifig3}
\end{figure}

In order to give the method a more stringent test, we use initial data with a significantly larger amplitude for the inhomogeneity of the scalar field.  Results for this simulation are presented in figures \ref{Hfig3} and \ref{phifig3}.  Here the results are somewhat different from those of the previous two simulations.  Though this simulation is run for twice as much time as the previous simulations, nonetheless even at the end of the simulation there are two distinct regions. In one region the scalar field has reached the bottom plateau of the potential and thus inflation has ended.  However in the other region the scalar field is still at the top plateau of the potential and it is evolving very slowly.  We can therefore expect inflation in this region to go on for an extended period of time.  Furthermore, the behavior of $H$ is becoming quite steep in the transiton region between the inflationary region and the region where inflation has ended.

\begin{figure}
\centering
\includegraphics[width=0.8\textwidth]{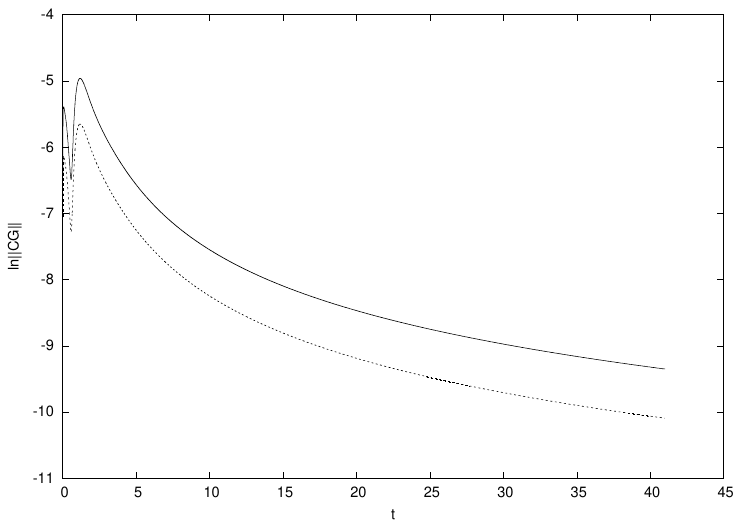}
\caption{$\ln (||{{\cal C}_G}|| )$ vs. $t$ at two different spatial resolutions.} 
\label{constraintfig}
\end{figure}

In cases in which gradients are steep, one must ensure that there are enough spatial points to provide adequate spatial resolution.  To address this issue, we present the results of a convergence test in figure (\ref{constraintfig}).  In this figure, we plot the natural logarithm of the $L_2$ norm of the constraint ${\cal C}_G$ of eqn. (\ref{cnstrG}) vs. time at two different resolutions.  The top curve is at the resolution of the simulations of figures (\ref{Hfig3}) and (\ref{phifig3}).  The bottom curve is with twice as many spatial points.  The results show that we are in the convergent regime and thus have adequate spatial resolution.

\section{Conclusion}
\label{Conclude}

Because the mechanism of inflation is local, inhomogeneous cosmologies remain inhomogeneous even when inflation occurs. This is because different regions may undergo different amounts of inflation and exit inflation at different times.  Thus a numerical cosmology method needs to be able to accurately evolve the spacetime for sufficiently long times even in the presence of ongoing and possiblely large inhomogeneities.  We have shown (in a proof of concept way) that mean curvature flow is a promising method for such robust long term evolution.  It would be interesting to do a comparison of robustness (using the same initial data) with the slicing methods used in refs. \cite{east1}-\cite{meannapaul}.

One limitation of our current simulations is the restriction to dependence on one spatial coordinate.  In particular this limitation does not allow us to treat the case where black holes form in an inhomogeneous inflating cosmology.  It would be interesting to perform simulations using our method in the case of no symmetry.  Since under mean curvature flow the mean curvature remains positive, we expect that in such a simulation the slices would slow down and essentially freeze in the black hole region. That is, we would expect a ``collapse of the lapse'' phenomenon similar to what occurs in maximal slicing.

\section*{Acknowledgments}

It is a pleasure to thank Frans Pretorius, Paul Steinhardt, and Anna Ijjas for helpful discussions.  This work was supported by NSF Grant  PHY-2102914.

\end{document}